\documentclass[namedreferences]{SolarPhysics}
\usepackage[optionalrh]{spr-sola-addons} 
\usepackage{graphicx}        
\usepackage{amssymb}        
\usepackage{color}           
\usepackage{url}             

\newcommand{\aap}{    {\it Astron. Astrophys.}}

\newcommand{\solphys}{{\it Solar Phys.}}

\begin{document}

\begin{article}

\begin{opening}

\title{A Technique for Removing Background Features in SECCHI--EUVI  He {\sc ii} 304 \AA\ Filtergrams: Application to the Filament Eruption of 22 May 2008}

\author{G.~\surname{Artzner}$^{1}$\sep
        S.~\surname{Gosain}$^{2}$\sep
        B.~\surname{Schmieder}$^{3}$
       }
\runningauthor{G. Artzner {\it et al.}}
\runningtitle{Background Feature Removal in STEREO Images}

   \institute{$^{1}$ Institut d'Astrophysique Spatiale B${\rm \hat{a} }$t. 121, 91405, Orsay, France
                     email: \url{Guy.Artzner@ias.u-psud.fr} \\
              $^{2}$ Udaipur Solar Observatory, P. Box 198, Dewali, Udaipur 313001, India
                     email: \url{sgosain@prl.res.in} \\
              $^{3}$ Observatoire de Paris, LESIA, 92190 Meudon, France
                     email: \url{Brigitte.Schmieder@obspm.fr} \\
             }

\begin{abstract}
 The STEREO mission has been  providing stereoscopic view of filament eruptions in EUV. The clearest view during a filament eruption is seen in He {\sc ii } 304 \AA\ observations. One of the main problems visualizing filament dynamics in He {\sc ii} 304  \AA\ is the strong background contrast due to surface features. We present a technique that removes background features and leaves behind only the filamentary structure, as seen by STEREO-A and B. The technique uses a pair of STEREO He {\sc ii} 304 \AA\ images observed simultaneously.  The STEREO-B image is geometrically  transformed to  STEREO-A view so that the background images appear similar. Filaments being elevated structures, {\it i.e.}, not lying on the same spherical surface as background features, do not appear similar in the transformed view. Thus, subtracting the two images cancels the background but leaves behind the filament structure.  We apply this technique to study the dynamics of the filament eruption event of 22 May 2008, which was observed by STEREO and followed by several ground-based observatories participating in the Joint Observing Programme (JOP 178).
\end{abstract}
\end{opening}

\section{Introduction}
     \label{S-Introduction}
Stereoscopic observations of  filament eruption in EUV
wavelengths are now routinely recorded by the SECCHI--EUVI
instrument onboard the STEREO spacecraft \cite{Kaiser08}. Using
these stereoscopic observations we can reconstruct the
three-dimensional geometry and the trajectory of erupting
filaments  \cite{Gissot08,Liewer09,Gosain09}. Of these EUV
observations, the He {\sc ii} 304 \AA\ observations are most
useful in tracing filaments because: {\it i)} the filament
spine appears much sharper and clearer \cite{Martin07,Joshi07},
and {\it ii)} the filament can be traced up to higher altitudes
compared to H$\alpha$ images \cite{Joshi07}.  For the quiet
Sun, He {\sc ii} 304 \AA\ corresponds to plasma temperatures of
$\approx 60\,-\,80 \times 10^{3}$ K. At these wavelengths, the
bright and dark  background features overlay network and
intra-network elements, respectively. The filaments are seen in
absorption as dark structures on the disk and are seen in
emission as prominence above the limb. During erupting phase,
the filaments seen on disk become diffuse and lose contrast and
are not very easy to distinguish against background features.
Studying the dynamics of erupting filaments is very important
to understand the global equilibrium of filaments and has
relevance for space weather prediction.

Here, we present a technique for  improving the contrast of the
filaments by  removing the background features as seen in He
{\sc ii} 304 \AA\ filtergrams. The technique uses a pair of
simultaneously observed STEREO-A and STEREO-B images. After
data reduction, the STEREO-B image is geometrically transformed
into the STEREO-A view, so the background images appear
similar.  Filaments being elevated structures, {\it i.e.} not
lying on same spherical surface as the background features, do
not appear similar in the transformed view. Thus, by taking
difference of the two images, the background features cancel
and only the elevated structures like filaments appear. We call
this the difference method. This difference method is valid
only if the filament is visible on the disk in both STEREO
images.  With this technique it is possible to detect faint
filament structures by removing the background features which
enhances the filament visibility. The larger separation between
the viewing angles is an advantage as the projections are very
different for elevated structures like filaments as compared to
surface features like plages. This results in a better contrast
in the difference image due to non-overlapping of elevated
structures in the transformed view.   Making movies out of such
difference images allows us to  visualize the dynamics of the
filaments in a much better way. We present, as an illustration,
an application of this technique to a filament disappearance or
disparition brusque (DB) event of 22 May 2008, when the
separation angle between STEREO satellites is about
52.4$^\circ$.

\inlinecite{Mouradian89} classified DBs into two categories,
one due to dynamic and the other one due to thermal
instability. In case of filament disappearance due to thermal
instability the filament reappears after the plasma has cooled
down. In the 22 May 2008 event the initial dense filament
became an  untwisting flux rope with multiple threads with a
fan-shaped structure which rose and disappeared one by one into
the corona. The plasma becomes optically so thin as the flux
rope rapidly expands in the corona that it remains no longer
visible.  So, although the filament is not dense enough to be
seen as erupting filament at larger distances from the Sun, we
can rule out thermal instability to be the cause of this event,
as the filament did not reform at this location even after
several days.

The key point here is that  for widely separated STEREO views
we could detect filament dynamics of this faint event by
applying the difference method and removing the background
features. The movie made out of these difference images show
the filament dynamics with much more clarity than the
traditional movies where the  background chromospheric features
are present. With the difference movie we make a reasonable
interpretation that this filament is eruptive, which is not so
obvious in normal movies.

Details of the technique of  geometric transformation are given
in Section 2. The case study of the 22 May 2008 filament
eruption is presented in Section 3. In Section 4 we discuss the
technique and results of the case study.

\section{Technique of Background Feature Removal}

\subsection{ Illustration of Method}
\begin{figure}    
\centerline{\includegraphics[width=0.8\textwidth,clip=]{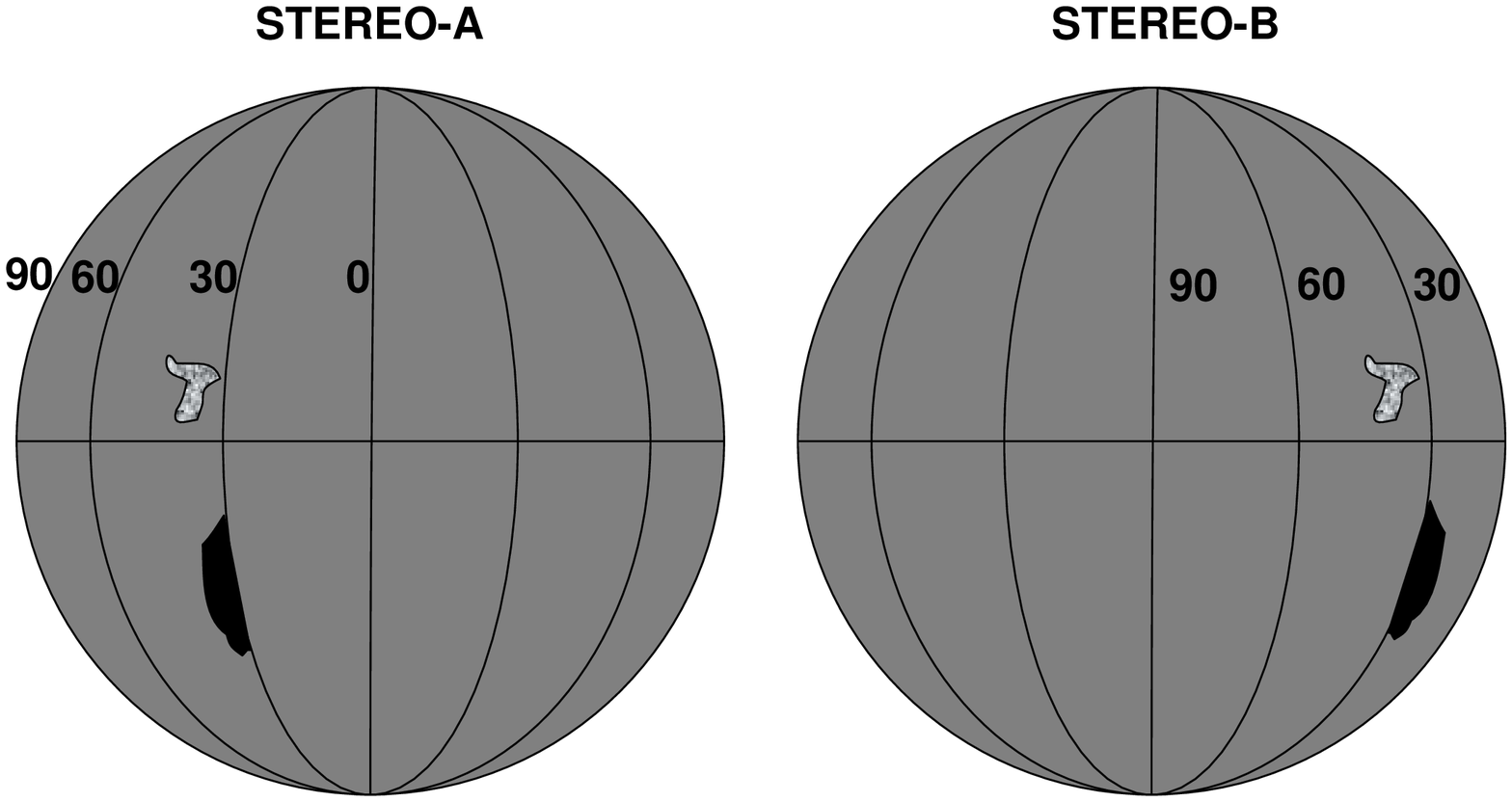}}
\vspace{0.2in}
\centerline{\includegraphics[width=0.8\textwidth,clip=]{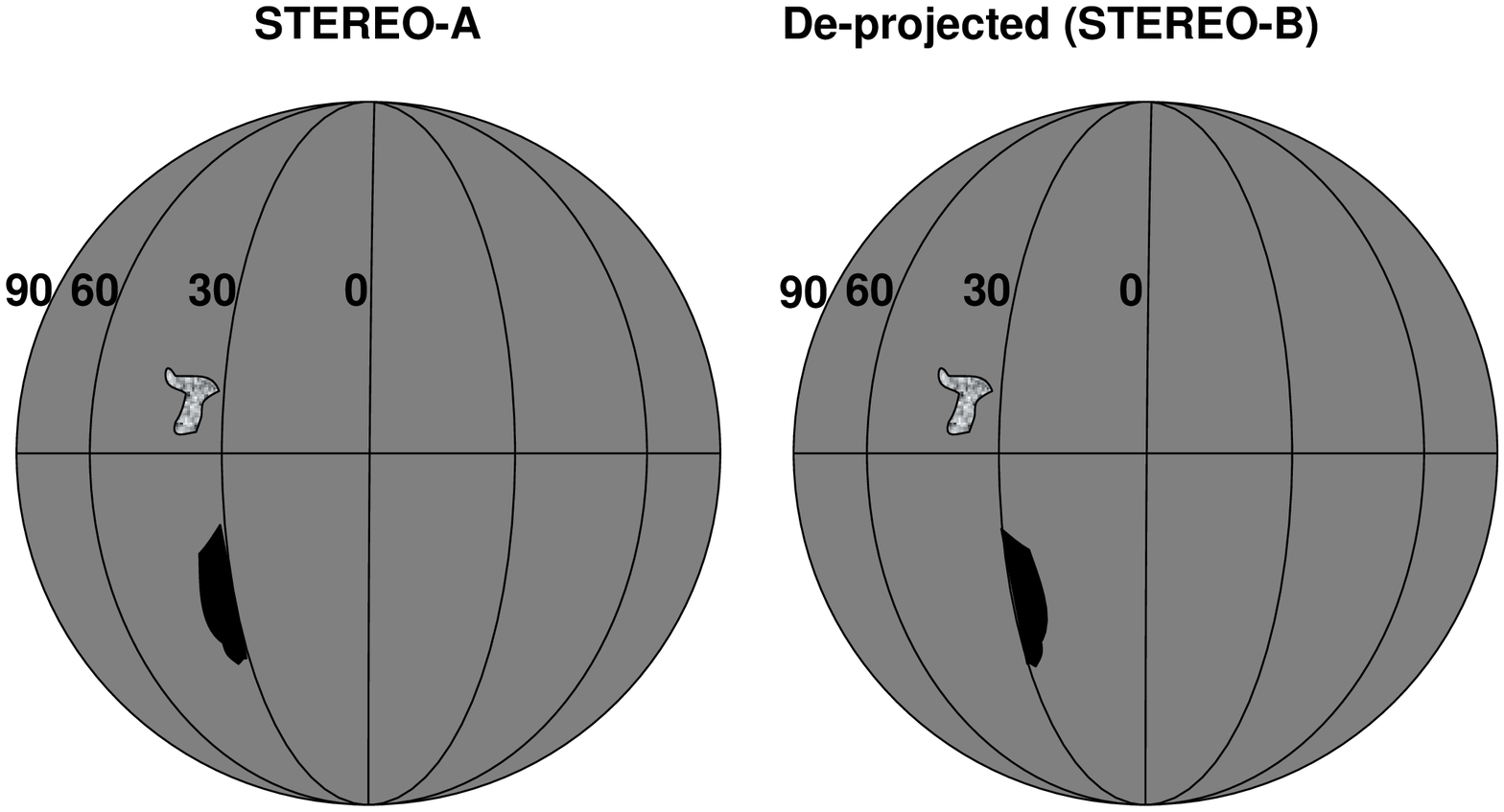}}
\vspace{0.2in}
\centerline{\includegraphics[width=0.4\textwidth,clip=]{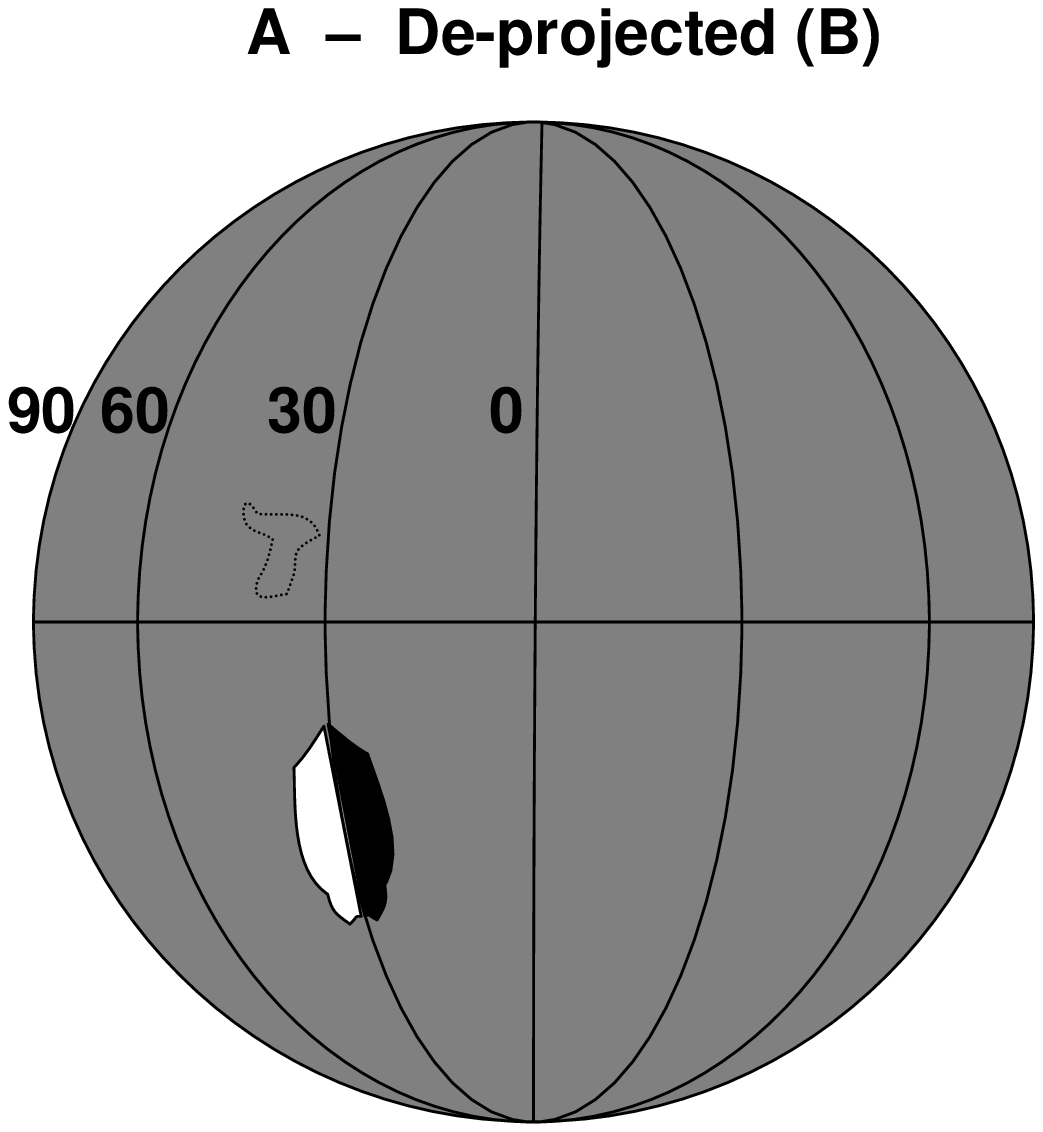}}
\caption{An illustration of the basic idea behind the method.
 Top:  The top panel illustrates a filament and a plage seen from
different angles by  STEREO A and B. The images are represented in epipolar view, with the grid representing the ``SECCHI Grid". While plage is a surface
feature, the filament is a three-dimensional structure in the solar
corona. The filament is assumed to be in a vertical
sheet. Hence its projection is seen differently from STEREO A and B
as compared to the plage.  Middle:  On de-projecting the
STEREO-B image to the reference frame of STEREO-A we see that unlike
 plage the filament cannot be reconciled due to different
projections.  Bottom: On taking the difference of STEREO-A image
and de-projected STEREO-B image we see that the surface features
(illustrated by a plage) cancels nicely while the filament leads to an imbalanced (black and white)
difference image. }
\label{fig:illustr1}
\end{figure}

Since filaments are three-dimensional structures with heights
extending up to a few 100 Mm, they are seen differently when
viewed from different angles, specially when the separation
between the viewing angles is large. An illustration of this
effect is shown in Figure~\ref{fig:illustr1}. The top panel
shows a filament and a solar surface feature {\it i.e.}, a
plage, observed from different angles by  STEREO A and B
satellites. Both images represent solar disk in the epipolar
view \cite{Inhester06}, {\it i.e.}, the central latitude
(equator) in both the images represents the epipolar line.
While plage is a background feature, the filament is a
three-dimensional structure extending into the solar corona.
Hence its projection is seen differently from STEREO-A and B as
compared to the background features. The background features
seen in He {\sc ii} 304 \AA\ images are assumed to be on a
sphere of radius R$\approx$ 696 Mm (photosphere) + 4 Mm
(chromosphere) = 700 Mm. On transforming the STEREO-B image to
the reference frame of STEREO-A, we see that unlike the plage,
the filament is seen differently due to different projections.
However, one must be careful because the plage is not observed
from the same point of view, so that transforming it using the
interpolation step may introduce some distortion effects.

Taking the difference of STEREO-A image and the transformed
STEREO-B image, we see that the surface feature (plage) cancels
nicely while the filament leads to imbalanced difference image.
This difference depends on: {\it i)} the location of the
filament on the solar disk with respect to the observer, {\it
i.e.,} the filament should be on the disk (for both STEREO A
and B views) despite the large angle between the twin
satellites, {\it ii)} the height of the filament, and {\it
iii)}  the inclination of the filament sheet with respect to
the local solar vertical direction.  In
Figure~\ref{fig:illustr1}, we have neglected the inclination of
the filament for simplification. The inclined filaments will in
general pose a difficulty in interpreting the structures,
specially if the spine is oriented along the line-of-sight,
hiding the filament structure below itself.

\subsection{STEREO Image Transformation }
The geometric illustration in Figure 2 presents the notations
used in the text below describing the method of transformation.
Let \url{hdra} and \url{hdrb} be the IDL structures of the
headers for the A and B images of a STEREO/SECCHI--EUVI pair.
The points in the physical space are noted with  ${\it O}$ as
the center of the solar sphere, ${\it A}$ and ${\it B}$ as the
position of STEREO-A and B respectively, ${\it M}$ as
intersection of the line-of-sight of the current point of the
STEREO-A image with a transparent reference sphere of arbitrary
radius ${\it R_{{\rm sol}}}$ concentric to the Sun. Relative to
the sphere: the center [o], the north pole [r] of the solar
Equator, the south pole [s] of the normal of the mission plane
are determined by the Sun and the STEREO spacecrafts. The
eastern, towards Earth, direction of the normal to the plane
determined by STEREO-A and the Sun's axis of rotation is
indicated by the point ${\it a}$  on the sphere. The eastern,
away from Earth, direction of the normal to the plane
determined by STEREO-B and the Sun's axis of rotation is
indicated by the point ${\it b}$  on the sphere.

\begin{figure}    
\centerline{\includegraphics[width=0.8\textwidth,clip=]{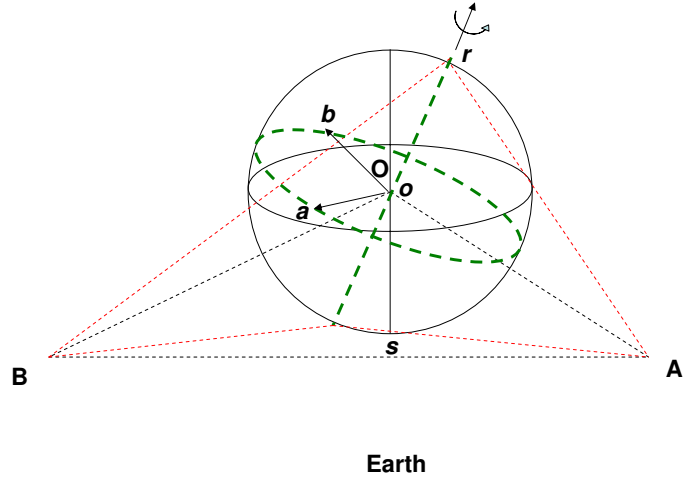}}
\caption{ Geometric illustration of the angles used for de-projecting the STEREO images.
 The dashed line and ellipse in green color show rotation axis of Sun and Equator respectively in ``Carrington grid" representation.
  The solid line and ellipse in black color show the epipolar axis and equator  respectively in ``SECCHI grid" representation. ``O" and ``o" are the
   same points, {\it i.e.}, the centre of Sun. $Oa$ is normal to plane $OrA$, and $Ob$ is normal to plane $OrB$, where $Or$ is rotation axis of the Sun.}
\label{fig:illustr2}
\end{figure}

The position of the south pole [s] normal to the mission plane
is given by the vector, ${\bf os = OA \times OB / |OA \times
OB|}$. The position of ${\it a}$ is  given by the vector ${\bf
oa=OA \times Or / |OA \times Or|}$ and the position of ${\it
b}$ is given by the vector ${\bf ob=OB \times Or / |OB \times
Or|}$.

The trace of the mission plane in each image is  the epipolar
plane. Both images are best compared when their epipolar plane
are aligned. The angle from polar north to the epipolar plane
is, $\mathrm{acos}(\bf{os} \cdot \bf{oa})$. The roll angle of
the satellite is given in the image
header as the CROTA parameter. The Chasles relation gives the amount of rotation, \\
$\mathrm{ -hdra.CROTA-90+ acos(\bf{os} \cdot \bf{oa})}$ \\
 and \\
$ \mathrm{-hdrb.CROTA-90+ acos(\bf{os} \cdot \bf{ob})}$ \\
to be applied to the raw images in order to obtain epipolar
plane parallel to the {\it x}-axis.

By analogy with  the Carrington grid, we then introduce the
notion of ``SECCHI Grid" of meridian circles  perpendicular to
and of latitude circles parallel with the mission  plane.  In
the figure ``SECCHI grid", the longitude range spans from
90$^\circ$ to the East to 90$^\circ$ to the West of the central
meridian, as seen from STEREO-A.  Homologous points have to be
located on the same latitude circle. When homologous points are
not located at the same relative position with respect to the
meridian great circles, a simple reasoning indicates where the
relevant structure is located, above or below the level of the
sphere upon which the reference grid is drawn.

\subsection{De-projection of STEREO-B image to STEREO-A}
Let ${\it i_{\rm a}}$ and ${\it j_{\rm a}}$ be the integer
coordinates of the current image point ${\it P_{\rm a}}$  of
the STEREO-A image. As a first step, these integer raw
coordinates are transformed into fractional centered
coordinates by using the  parameters ${\it CRPX_{\rm i}}$  in
the header, and further rotated to the mission-plane system
with the rotation as determined above with the Chasles
rotation. Taking into account the values of two distances, the
radius of the reference sphere and the Sun to spacecraft
distance, and the scale factors ${\it CDELT_{\rm j}}$ in the
header of the STEREO-A image converting pixel numbers into
angles, the set of two  fractional coordinates pertaining to
${\it P_{\rm a}}$ allows us to compute by classical geometry
the $x$, $y$, $z$ coordinates of the physical point ${\it M}$
corresponding to ${\it P_{\rm a}}$. The ``SECCHI-A" latitude
and longitude of the point ${\it M}$ are then computed
respectively as $\mathrm{acos}(y/\sqrt{x^2+y^2})$ and
$\mathrm{atan}(z,x)$, in order to possibly add a ``SECCHI grid"
to the images. The ``SECCHI-B" $z$-coordinate of the point
${\it M}$ is, by definition of the SECCHI system, equal to its
``SECCHI-A" value. The ``SECCHI-A" to ``SECCHI-B"
transformation for the $x$ and $y$ coordinates is obtained by
the rotation of the angle between ${\it A}$ and ${\it B}$, as
seen from the Sun. Using again  the values of two distances,
the radius of the reference sphere and the Sun to STEREO-B and
the scale factor as indicated by the parameters ${\it
CDELT_{\rm j}}$ in the header of the STEREO-B image, we compute
the fractional coordinates of the homologous point ${\it P_{\rm
b}}$ in the STEREO-B image. It should be noted that in this
procedure we compute exactly the coordinates of the homologous
point in the STEREO-B image related to the current point in the
STEREO-A image. We do not use the approximations used when
rotating images. One approximation that we use is  a bi-cubic
spline function, when interpolating  the measured intensity in
the STEREO-B image  relative to the physical point ${\it M}$.

\begin{figure}    
\centerline{\includegraphics[width=1.\textwidth,clip=]{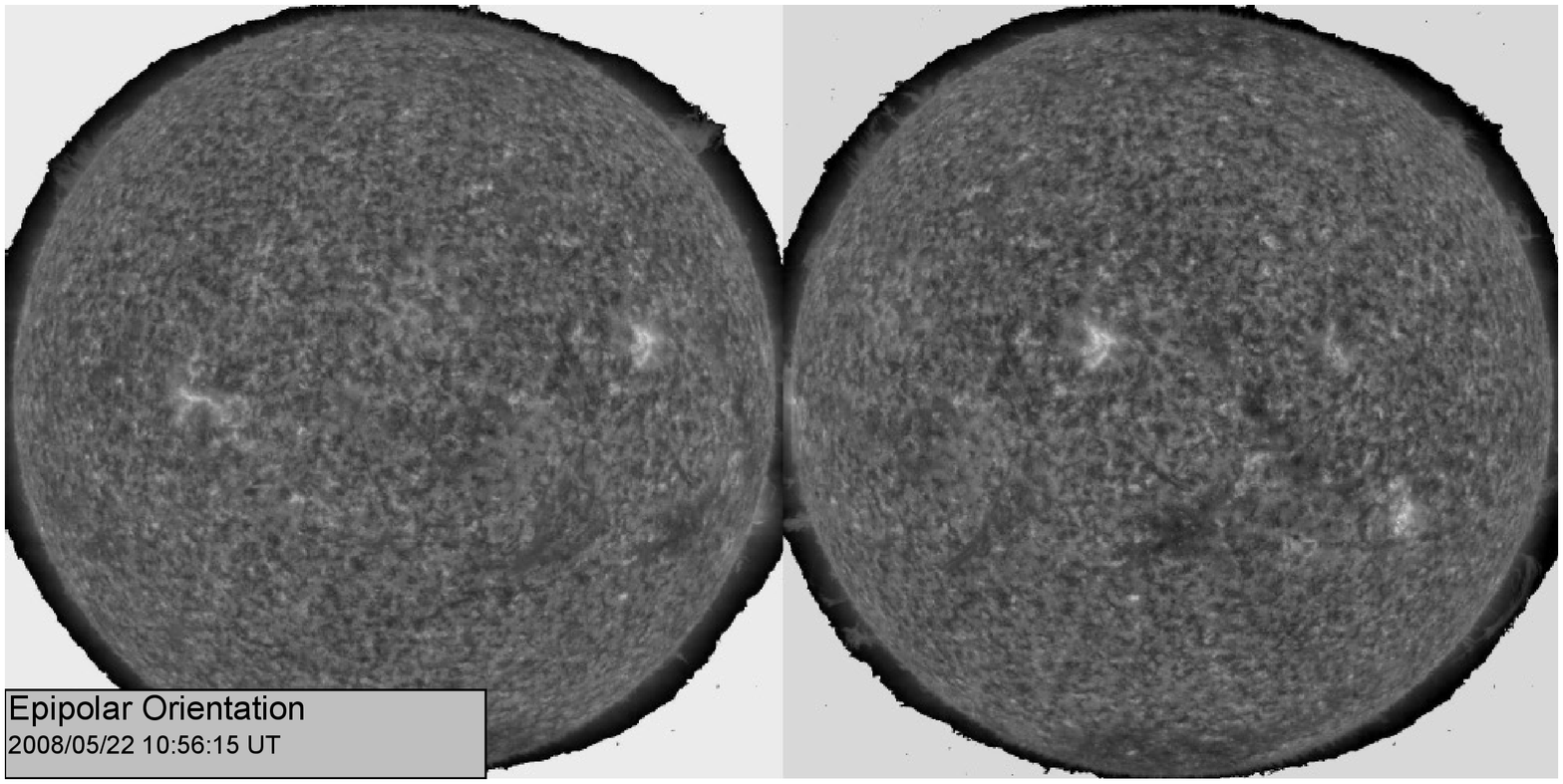}}
\centerline{\includegraphics[width=1.\textwidth,clip=]{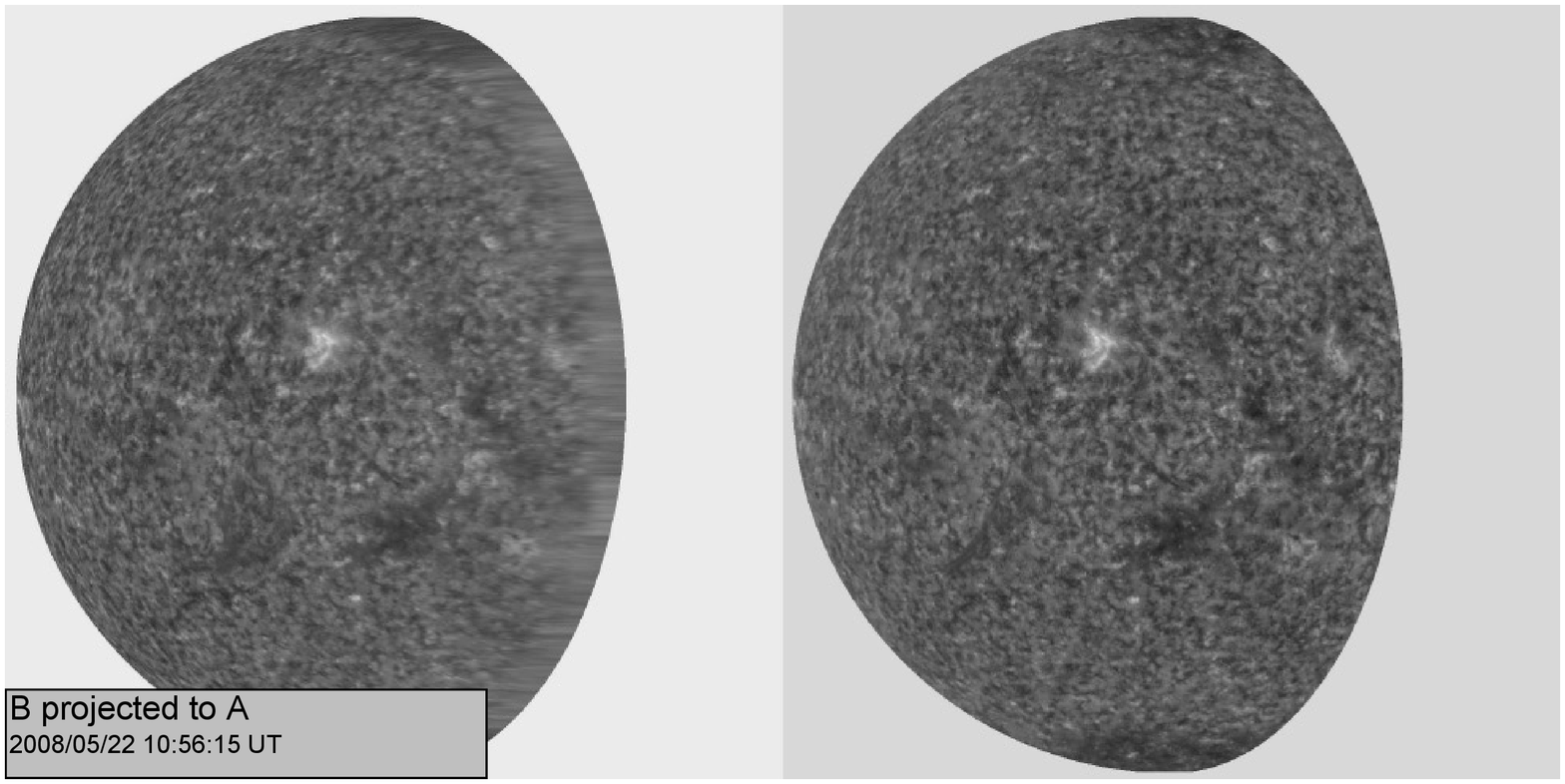}}
\centerline{\includegraphics[width=0.45\textwidth,clip=]{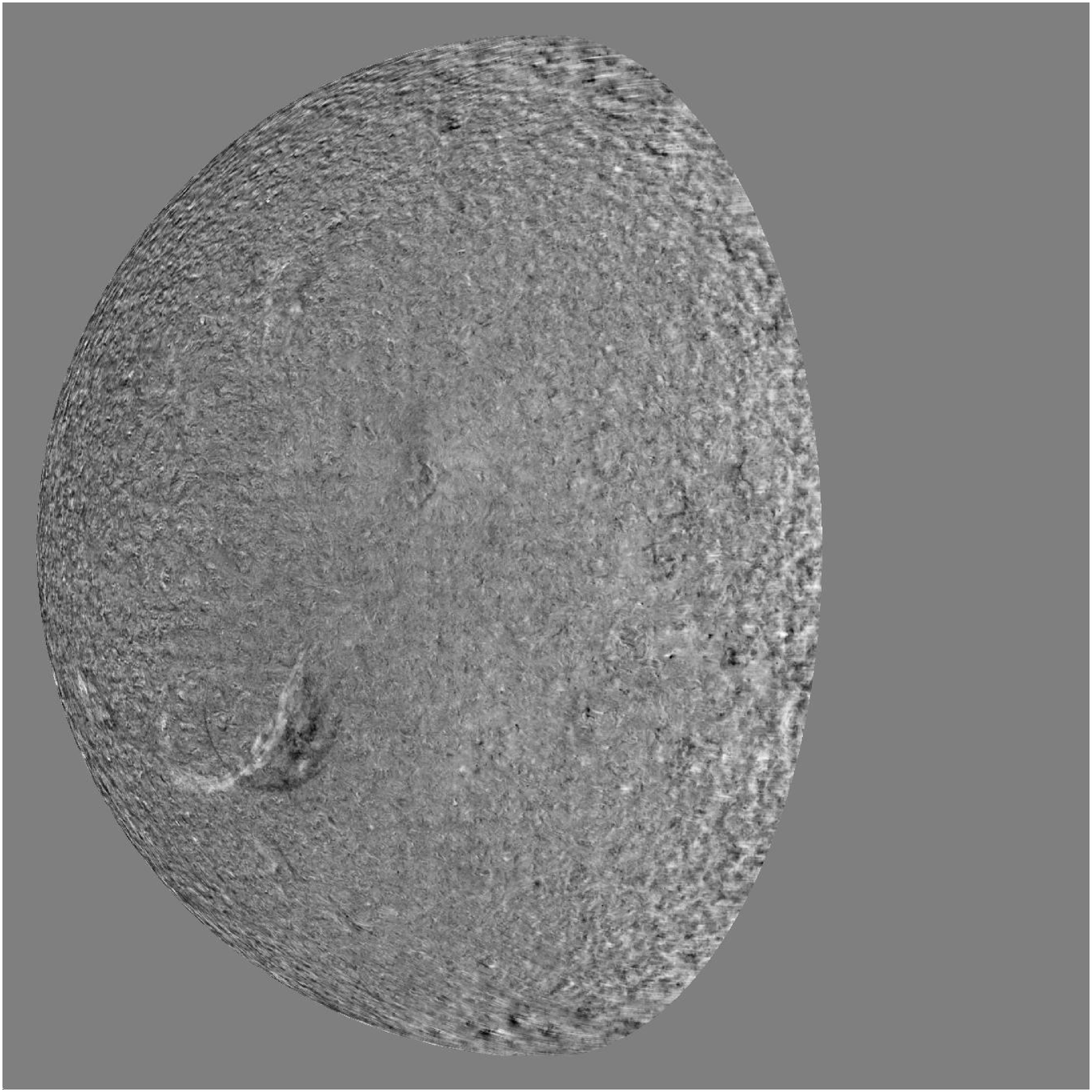}}
\caption{ Top panel shows STEREO A, B images in epipolar orientation after the raw images have been centered and re-scaled.
 The middle panel shows the STEREO B image projected to STEREO A. Bottom panel shows the difference of the two images in the middle panel.
 It can be noticed that only the filament and other elevated structures remain as black--white difference features
  while  background features are completely removed, particularly the white plage. }
\label{fig:illustr3}
\end{figure}

\section{ Application to STEREO Data}
\subsection{ Basic Steps }
The basic steps required are as follows:
\begin{itemize}
\item \url{SECCHI_PREP} data reduction
\item Centering of images
\item Re-scaling the images for common size of solar disk.
\item Correction for the roll orientation. Constrained by the criteria that common or homologous features lie on the epipolar line.
\item Transformation of STEREO-B to A.
\item Taking the difference image.
\end{itemize}

These steps are demonstrated on the real observations in Figure
3. These STEREO observations are of a filament eruption on 22
May 2008 at 10:56 UT observed in He {\sc ii} 304 \AA. The
caption of Figure 3 explains the different steps applied to
these observations to yield a difference image.

\subsection{Global View of Filament Eruption on 22 May 2008}
\begin{figure}    
\centerline{\includegraphics[width=1.\textwidth,clip=]{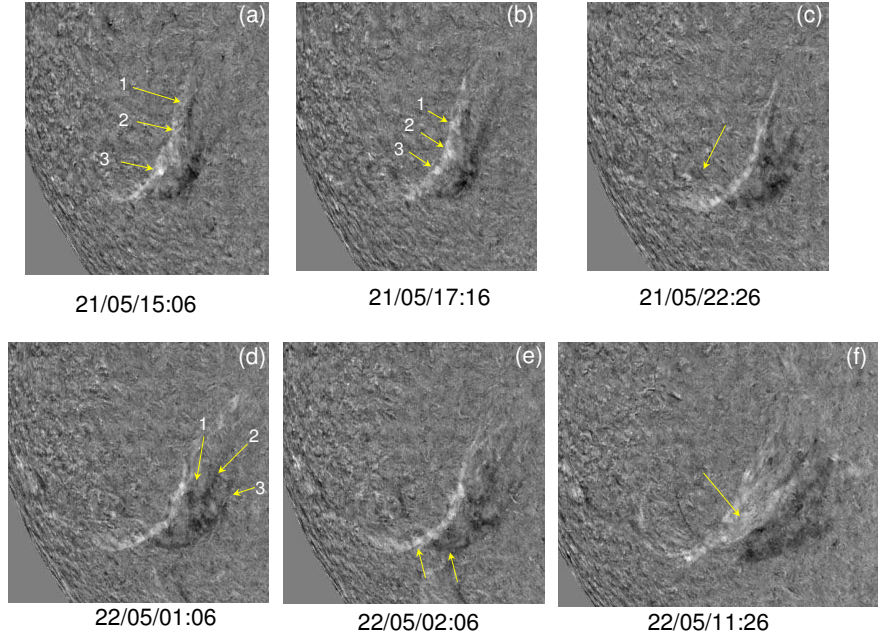}}
\caption{ Difference images obtained with He {\sc ii} 304 \AA\ filtergrams (STEREO A
is white, STEREO B is black). The description of the panels is as follows: (a) The filament is helicoidal with three turns shown by arrows,
 numbered as 1, 2, and 3. (b) The three branches (blobs), numbered as 1, 2, and 3, are shown by arrows. (c) A thin filament spine is visible at the location marked by arrow. (d) Filament seems to split into three branches along its length, numbered as 1, 2 and 3. (e) Apparent twist  in the filament is represented by arrows. (f) The filament is in erupting phase, the STEREO-A view (white part) is now expanded with a fan shape. A movie of all the difference images (available online as supplementary material) shows oscillating fine threads during this time.   }
\label{fig:diffimg}
\end{figure}

The filament eruption of 22 May 2008 has been previously
studied by \inlinecite{Gosain09}. The height of the filament
was derived using the triangulation method \url{SCC_MEASURE}
\cite{Thompson06} for one set of observations during the
eruption phase at 10:56 UT on 22 May. The filament spine
reached an altitude of $\approx$ 180 Mm. Here, we  use the
difference images to derive the qualitative information about
the filament as it evolved and finally erupted. To this end, we
made a movie of the difference images which is made available
as electronic supplementary material (\url{movie1.mpg}). The
time sequence of such difference images are useful to study the
evolution of filament. In Figure~\ref{fig:diffimg}, we show the
difference images created using STEREO-A and B observations of
22 May during different phases of the eruption. We summarize
the qualitative information about the filament evolution, using
these difference images, in  Table 1. Initially, the erupting
section of the filament has a helical shape with an estimated
three turns. Branching into three parts appear on one end
(upper end) of the filament. Subsequently, we see  thin long
filamentary structures with a slow rising motion. Later, the
spine disappears and only the three branches remain visible,
which erupt with a oscillating motion.  All the threads are
very thin and will not be visible with a low spatial resolution
telescope. The filament structure was inflating before rising
and erupting, and the material cools to escape along field
lines until it is no more distinguishable because of its weak
contrast. The filament eruption was rather slow, with
successive eruptive phases concerning different parts of the
filament, and lasted for about 24 hours, disappearing
completely by about 18:00 UT on 22 May 2008.  After viewing the
difference movie (\url{movie1.mpg}) we get an impression that
it is a slow eruption with top threads disappearing slowly. If
it was a thermal DB it would perhaps reappear as the plasma
cools down, we do not see such reappearance for next few days.

While the structure of the filament can be studied with the
difference images, we can also use difference images to detect
filaments over the solar disk.  By clipping the difference
signals for the filament, {\it i.e.} putting 1 or 0 for
negative or positive difference values respectively above
certain threshold, we made movies of the filament
disappearance. This movie is useful to  identify the  filaments
on the disk, specially when an automated identification is
required. The movie is included as electronic supplementary
material (\url{movie2.mpg}).

\begin{table}
\caption{Sequence of events in He {\sc ii} 304 \AA\
observations} \label{tab:2}
\begin{tabular}{lcr}
\hline
Observation &  Date  &  Time (UT) \\
\hline
1. Helicoidally shape with three turns&             21 May & 15:06 \\
2. Three branches are visible      &            21 May & 17:16\\
3. Thin filament  at the latitude -20$^{\circ}$ &21 May & 22:26\\
4. Filament  changing  with three&22 May &  01:06\\
~~~ branches instead of a spine&& \\
5. Twist                                            &22 May& 02:06 \\
6. Fan and oscillating threads&     22 May & 11:26 \\
7. Complete disappearance & 22 May & 18:00 \\
\hline
\end{tabular}
\end{table}

\subsection{ Partial Filament Eruption }
\begin{figure}    
\centerline{\includegraphics[width=1.0\textwidth,clip=]{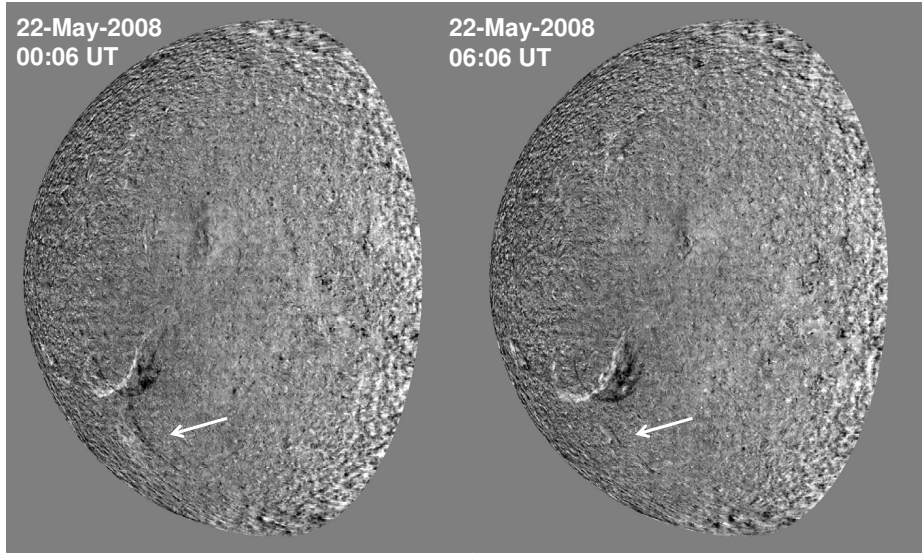}}
\caption{ The difference images during 00:06 UT and 06:06 UT shows the partial filament (marked by white arrows) disappearance.}
\label{fig:context}
\end{figure}

With the help of difference images we find that there is
another faint filament located close (southwards) to the
filament discussed above, which erupts earlier during 02:00 UT
to 07:00 UT on 22 May 2008. The two difference images displayed
in the upper panel of Figure~\ref{fig:context}, show the
disappearance of this faint filament (marked by white arrows).
From the movie of difference images it appears that the two
filaments are connected closely. So, we can consider this faint
filament eruption as a case of partial filament eruption. The
part of the filament which disappeared is represented by the
white arrow in the mosaic. A CME was detected by the STEREO-A
COR2, {\it Solar Mass-Ejection Imager} (SMEI) and {\it Large
Angle and Spectrometric Coronagraph} (LASCO) C2 instruments
around this time.  A movie of STEREO-A COR2 observations of the
CME is made available as electronic supplementary material
(\url{movie3.mpg}).  However, the association of the CME with
this partial filament eruption could not be clearly established
due to another off-limb prominence eruption nearby on 21 May
2008. It could also happen that the CME associated with partial
filament eruption was too faint to  be detected by these
instruments.

Filament eruptions and associated CMEs are commonly interpreted
in terms of helical magnetic flux rope structure
\cite{Amari00}. Whether, these flux ropes exist prior to
eruption or  they are formed during eruption, is still a
controversy. An alternative model, suggested by
\inlinecite{Gibson06}, is known as the ``partially-expelled
flux rope" (PEFR) model. In this model, a flux rope exists
prior to the the CME and as it erupts it reconnects internally
and with the surrounding fields and breaks into two flux ropes.
One portion of the rope is expelled as a CME while the other
flux rope remains attached. Several partial filament eruptions
were studied by \inlinecite{Tripathi09} and were found to be
consistent with the PEFR model. In our observations, the faint
filament eruption event could be a partially erupting flux rope
with either no CME at all or no CME detected. The other
remaining part of the flux rope, {\it i.e.}, the filament
discussed in previous section, erupts 8 to 12 hours later. The
second filament eruption could be due to destabilization caused
by the partial filament eruption.


\section{Discussion and Conclusions}
We describe a method that helps to study filament eruptions
using He {\sc ii} 304 \AA\ observations from STEREO. The
elevated structures are seen better if the separation angle is
large because the difference signal is enhanced as there is no
overlapping of structures. This method has the potential to be
used to detect not only the filaments but also other elevated
structures like plumes or jets. This could be very useful in
studies of  transient events.

In the present paper, the method has been applied to real
observations of a filament eruption on 22 May 2008, when the
separation angle of the two STEREO satellites was equal to
52.4$^\circ$. At such large separation angles the traditional
stereoscopic techniques, like triangulation method as
identifying common features in both images is difficult. Hence,
we need to use alternative techniques like the difference
method to extract the three dimensional information about the
filament. Here, a qualitative study of the eruption event is
done with the difference images, which gives a global view of
the event. The movie of the difference images
(\url{movie1.mpg}) shows clearly the evolution and dynamics of
the erupting filament,  which is not so clear in conventional
movies made of intensity pictures.  Further, a faint CME was
observed by STEREO-A COR2 on 22 May 2008, during 02:00 UT to
07:00 UT.  However, the association of this CME with this
partial filament eruption is ambiguous because of another
filament eruption on 21 May 2008 which occurred nearby off the
limb. This faint filamentary structure could be seen clearly in
the ``clean" images. Also, this event concerns a large region
on the Sun (about a quadrant of the visible solar disk), which
becomes evident in the movie of difference images of He {\sc
ii} 304 \AA. Also, with proper thresholds, the method could be
used to make automatic detection of filaments and transient
events. Thus, this image processing technique helps in the
studies of solar eruptive phenomena by isolating filament
features  from background chromospheric features. Also, with
the launch of {\it Solar Dynamics Observatory} (SDO) there is a
possibility of applying these techniques to STEREO-A--SDO,
STEREO-B--SDO pairs.

\begin{acks}
SG acknowledges CEFIPRA funding for his visit to the
Observatoire de Paris, Meudon, France under its project No.
3704-1. The STEREO/SECCHI data are produced by a consortium of
RAL (UK), NRL (USA), LMSAL (USA), GSFC (USA), MPS (Germany),
CSL (Belgium), IOTA (France), and IAS (France).
\end{acks}

\bibliographystyle{spr-mp-sola}
%

\end{article}

\end{document}